\documentclass[pageno]{jpaper}

\pdfoutput=1

\usepackage{amsmath,amssymb,amsfonts}
\usepackage{algorithmic}
\usepackage{graphicx}
\usepackage{textcomp}
\usepackage{xcolor}
\usepackage{fancyhdr}
\usepackage[normalem]{ulem} 
\usepackage{xspace}
\usepackage{scalerel} 
\usepackage{hhline}
\usepackage{upgreek}
\usepackage{footnote}
\usepackage{enumitem}
\usepackage{graphicx}
\usepackage[bottom]{footmisc}
\usepackage{hyperref}
\usepackage{textcomp}
\usepackage{array}
\usepackage{multicol}
\usepackage{multirow}
\usepackage{courier}
\usepackage{tikz} 
\usepackage{flushend}
\PassOptionsToPackage{hyphens}{url}\usepackage{hyperref}
\makeatletter
\g@addto@macro{\UrlBreaks}{\UrlOrds}
\makeatother

\newcolumntype{?}{!{\vrule width 1pt}}
\newcolumntype{P}[1]{>{\centering\arraybackslash}p{#1}}

\newcommand{\prr}{{PR$^{2}$}\xspace}
\newcommand{\arr}{{AR$^{2}$}\xspace}

\newcommand{\vth}{$\text{V}_\text{TH}$\xspace} 
\newcommand{\vref}[1]{$\text{V}_{\text{REF}#1}$\xspace} 

\newcommand{\vopt}{$\text{V}_{\text{OPT}}$\xspace}
\newcommand{\tr}{\texttt{tR}\xspace} 

\newcommand{\cread}{\texttt{CACHE}~\texttt{READ}\xspace}

\begin{document}

\title{Reducing Solid-State Drive Read Latency by Optimizing Read-Retry\\
\textbf{\textit{Extended Abstract}}}

\newcommand{\affilETH}{$^1$}
\newcommand{\affilSNU}{$^2$}
\newcommand{\affilKNU}{$^3$}

\author{
    \vspace{10pt}\\
    \scalebox{0.95}{
        {Jisung Park\affilETH}\quad%
        {Myungsuk Kim$^{2, 3}$}\quad%
        {Myoungjun Chun\affilSNU}\quad%
        {Lois Orosa\affilETH}\quad%
        {Jihong Kim\affilSNU}\quad%
        {Onur Mutlu\affilETH}\quad%
    }\\%
    \vspace{15pt}%
    {\it \affilETH ETH Z{\"u}rich  \quad  \affilSNU Seoul National University \quad {\affilKNU}Kyungpook National University}
    \vspace{-2.5em}%
}

\date{}

\maketitle

\thispagestyle{empty}

\section{Motivation}
\label{sec:motivation}

This work tackles the performance degradation of modern NAND flash-based SSDs due to a large number of \emph{read-retry} operations essential to ensuring the reliability of stored data.
While 3D NAND technology and multi-level cell (MLC) techniques enable continuous increase of storage density, they also negatively affect the reliability of modern NAND flash chips.
NAND flash memory stores data as the \emph{threshold voltage (\vth)} of each flash cell, which depends on the amount of charge in the cell.
New cell designs and organizations in 3D NAND flash memory cause a flash cell to more easily leak its charge~\cite{cai-arxiv-2017, cai-insidessd-2018, luo-hpca-2018, luo-sigmetrics-2018}.
In addition, MLC technology significantly reduces the margin between different \vth levels to store multiple bits in a single cell.
Consequently, the \vth level of a 3D NAND flash cell with advanced MLC techniques (e.g., triple-level cell (TLC)~\cite{kang-iccss-2016} or quad-level cell (QLC)~\cite{huh-iccss-2020, kim-isscc-2020}) can quickly shift beyond the read-reference voltage \vref{} after programming, which results in an error when reading the cell.

To provide reliability guarantees for stored data, a modern SSD commonly adopts two main approaches.
First, a modern SSD employs a strong \emph{error-correcting code (ECC)} that can detect and correct several tens of raw bit errors (e.g., 72 bits per 1-KiB codeword~\cite{micron-flyer-2016}).
Second, when ECC fails to correct all bit errors, the SSD controller performs a \emph{read-retry operation}~\cite{cai-date-2012} that reads the erroneous page again with \emph{slightly-adjusted} \vref{} values. 
Since bit errors occur due to shift of the \vth levels of flash cells beyond the \vref{} values, sensing the cells with appropriately-shifted \vref{} values can greatly reduce the number of raw bit errors~\cite{cai-procieee-2017, cai-arxiv-2017, cai-insidessd-2018, cai-hpca-2017, cai-date-2013, cai-dsn-2015, cai-hpca-2015, cai-iccd-2013, cai-inteltechj-2013, cai-sigmetrics-2014, fukami-dfrwseu-2017, luo-ieeejsac-2016, luo-hpca-2018, luo-sigmetrics-2018, shim-micro-2019}. 

Even though read-retry is essential to ensuring the reliability of modern NAND flash memory, it comes at the cost of significant performance degradation.
A read-retry operation \emph{repeats} a retry step that reads the target page while adjusting \vref, until it finds a \vref{} value that allows the page's raw bit-error rate (RBER) to be lower than the ECC correction capability.
Recent work~\cite{shim-micro-2019} shows that a modern SSD with long retention ages (i.e., how long data is stored) and high program/erase (P/E) cycles (i.e., how many program/erase operations are performed) suffers from a large number of read-retry operations, which in turn increases the read latency linearly with the number of retry steps.
Our experimental characterization using 160 real 3D TLC NAND flash chips, in this work, shows that a read frequently incurs \emph{multiple} retry steps even under modest operating conditions (e.g., on average 4.5 retry steps under a 3-month data retention age at \emph{zero} P/E cycles, i.e., at the beginning of SSD lifetime).

Considering that 1) read-retry operations would occur even more frequently in newer NAND flash memory, and 2) many key applications in modern computing systems (e.g., key-value stores and graph analytics) require high read performance on storage devices, it is important to minimize the performance overhead of read-retry operations.

\section{Limitations of the State of the Art}
\label{sec:limitations}

To mitigate the performance overhead of read-retry operations, prior works~\cite{cai-hpca-2015, cai-iccd-2013, luo-ieeejsac-2016, luo-sigmetrics-2018, shim-micro-2019} propose to keep track of pre-optimized \vref{} values for each page to use them for future read requests.
For example, Shim et al.~\cite{shim-micro-2019} propose to read a page using \vref{} values that have been recently used for a read-retry operation on other pages exhibiting similar error characteristics with the page to read.
By starting a read (and retry) operation with the \vref{} values close to the optimal read-reference voltage (\vopt) values, their proposal significantly reduces the number of retry steps in modern NAND flash-based SSD.

Although prior techniques are effective at reducing the number of retry steps on an erroneous page, read-retry is a fundamental problem \emph{hard to completely avoid} in modern SSDs.
For example, the state-of-the-art technique described above can reduce about 70\% of retry steps, but \emph{every read} incurs at least three retry steps in an aged SSD~\cite{shim-micro-2019}.
This is because, in modern NAND flash memory, the \vth levels of flash cells change  quickly and significantly over time, which makes it extremely difficult to identify the exact \vref{} values that can avoid read-retry before reading the target page.

\section{Key Insights}
\label{sec:key-insights}

We identify new opportunities to reduce the read-retry latency by exploiting two advanced features in modern SSDs: 1) the \emph{\cread command}~\cite{leong-uspatent-2008, macronix-technote-2013, micron-technote-2004} and 2) \emph{strong ECC engine}.
First, we find that it is possible to reduce the total execution time of a read-retry operation using the \cread command that allows a NAND flash chip to perform consecutive reads in a pipelined manner.
Since each retry step is effectively the same as a regular page read, the \cread also enables concurrent execution of consecutive retry steps. 

Second, we find that a large ECC-capability margin exists in the final retry step.
This may sound contradictory as a read-retry occurs only when the page's RBER exceeds the ECC capability, i.e., when there is no ECC-capability margin.
However, when a read-retry operation succeeds, the page is eventually read \emph{without} any uncorrectable error, which means that there always exists \emph{a positive} ECC-capability margin in the final retry step.
We hypothesize that the ECC-capability margin is large due to two reasons.
First, a modern SSD uses a \emph{strong} ECC that can correct several tens of raw bit errors in a codeword.
Second, in the final retry step, the page can be read by using \emph{near-optimal} \vref{} values that drastically decrease the page's RBER.
If we can leverage the large ECC-capability margin to reduce the \emph{page-sensing latency \tr}, we can optimize the latency of \emph{every} retry step.
Doing so can allow not only the final retry step to quickly read the page without uncorrectable errors but also the earlier retry steps (which would fail anyway with the default \tr) to be finished more quickly.
To validate our hypothesis, we characterize 1) the ECC-capability margin in each retry step and 2) the impact of reducing \tr on the page's RBER, using 160 real 3D TLC NAND flash chips.
The results show that we can safely reduce \tr of each retry step by 25\% even under the worst operating conditions prescribed by manufacturers (e.g., a 1-year data retention age~\cite{cox-fms-2018} at 1.5K P/E cycles~\cite{micron-flyer-2016}).

The optimization opportunities that we identify enable new techniques that reduce
\emph{the latency of each retry step without increasing the number of retry steps}.
Such techniques can effectively complement existing techniques~\cite{cai-hpca-2015, cai-iccd-2013, luo-ieeejsac-2016, luo-sigmetrics-2018, shim-micro-2019} that aim to reduce the \emph{number} of retry steps on an erroneous page.

\section{Main Artifacts}
\label{sec:main-artifacts}

We develop two new read-retry mechanisms that effectively reduce the read-retry latency.
First, we propose \underline{P}ipelined \underline{R}ead \underline{R}etry (\prr) that performs consecutive retry steps in a pipelined manner using the \cread command.
Unlike the regular read-retry mechanism that starts a retry step \emph{after} finishing the previous step, \prr performs page sensing of a retry step during data transfer of the previous step, which removes data transfer and ECC decoding from the critical path of a read-retry operation, reducing the latency of a retry step by 28.5\%.

Second, we introduce \underline{A}daptive \underline{R}ead \underline{R}etry (\arr) that performs each retry step with reduced page-sensing latency (\tr), leading to a further 25\% latency reduction even under the worst operating conditions.
Since reducing \tr inevitably increases the read page's RBER, an excessive \tr reduction can potentially cause the final retry step to fail to read the page without uncorrectable errors.
This, in turn, introduces one or more additional retry steps, which could increase the overall read latency.
To avoid increasing the number of retry steps, \arr uses the best \tr value for a certain operating condition that we find via extensive and rigorous characterization of 160 real 3D NAND flash chips.

Our two techniques require only small modifications to the SSD controller or firmware but no change to underlying NAND flash chips.
This makes our techniques easy to integrate into an SSD along with existing read-retry mitigation techniques that aim to reduce the number of retry steps.

We evaluate our techniques using MQSim~\cite{mqsim-git, tavakkol-fast-2018}, an open-source multi-queue SSD simulator.
We extend MQSim to simulate more realistic read-retry characteristics in modern SSDs based on our real-device characterization results.
We also evaluate the performance improvement of our techniques when combined with a state-of-the-art technique~\cite{shim-micro-2019}.
We use twelve real-world workloads with different I/O characteristics while varying the data retention age and P/E-cycle count.

\section{Key Results and Contributions}
\label{sec:key-contributions}

Our main evaluation results show that \prr and \arr, when combined, significantly improve the SSD response time, by up to 50.8\% (35.2\% on average) over a high-end SSD.
Compared to a state-of-the-art baseline~\cite{shim-micro-2019}, our proposal further reduces SSD response time by up to 31.5\% (17\% on average) in read-dominant workloads.

This paper makes the following key contributions:
\begin{itemize}[leftmargin=*]
    \item To our knowledge, this work is the first to identify new opportunities to reduce the latency of each retry step by exploiting advanced architectural features in modern SSDs.
    \item Through extensive and rigorous characterization of 160 real 3D TLC NAND flash chips, we make three new observations on modern NAND flash memory.
    First, a read-retry operation with multiple retry steps frequently occurs even under modest operating conditions.
    Second, when a read-retry occurs, there is a large ECC-capability margin in the final retry step even under the worst operating conditions.
    Third, there is substantial margin in read-timing parameters, which enables safe reduction of the read-retry latency.
    \item Based on our findings and characterization results, we propose two new techniques, \prr and \arr, which effectively reduce the latency of each retry step, thereby reducing overall read latency and thus improving application performance.
    Our techniques require only very small changes to the SSD controller or firmware.
    By reducing the latency of each retry step while keeping the same number of retry steps during a flash read, our proposal effectively complements existing techniques~\cite{cai-hpca-2015, cai-iccd-2013, luo-ieeejsac-2016, luo-sigmetrics-2018, shim-micro-2019} that aim to reduce the number of retry steps, as we empirically demonstrate in the paper.
\end{itemize}

\vspace{0.5em}
\noindent
\textbf{Why ASPLOS?}
Our work emphasizes the synergy between two fundamental aspects of storage systems: 1) firmware (i.e., system software) and 2) architecture.
Read-retry is an essential mechanism in SSD firmware to ensure the reliability of storage systems, but it can significantly degrade SSD I/O performance that is critical to data-intensive applications.
Through extensive real-device characterizations, we introduce new opportunities to significantly reduce read-retry latency by exploiting advanced architectural features widely adopted in modern SSDs.
Therefore, this work emphasizes the importance and effectiveness of optimizations based on comprehensive understanding of the storage firmware, architecture, and device characteristics.

\vspace{0.5em}
\noindent
\textbf{Citation for Most Influential Paper Award.}
This paper proposes new techniques to optimize the read-retry mechanism, which is essential to ensuring the reliability of modern NAND flash-based SSDs at the expense of significant latency overhead.
This work is the first to demonstrate that the large reliability margin in modern SSDs can be used to improve the read latency, which has impacted many real SSD designs and inspired many creative follow-on works to achieve high I/O performance by better exploiting the performance-reliability trade-off.

\bibliographystyle{plain}
\bibliography{reference}

\begin{thebibliography}{10}

\bibitem{mqsim-git}
{MQSim GitHub} repository.
\newblock \url{https://github.com/CMU-SAFARI/MQSim}.

\bibitem{cai-procieee-2017}
Yu~Cai, Saugata Ghose, Erich~F. Haratsch, Yixin Luo, and Onur Mutlu.
\newblock Error characterization, mitigation, and recovery in
  flash-memory-based solid-state drives.
\newblock {\em Proc. IEEE}, 2017.

\bibitem{cai-arxiv-2017}
Yu~Cai, Saugata Ghose, Erich~F. Haratsch, Yixin Luo, and Onur Mutlu.
\newblock Errors in flash-memory-based solid-state drives: Analysis,
  mitigation, and recovery.
\newblock {\em arXiv}, 2017.

\bibitem{cai-insidessd-2018}
Yu~Cai, Saugata Ghose, Erich~F. Haratsch, Yixin Luo, and Onur Mutlu.
\newblock Reliability issues in flash-memory-based solid-state drives:
  Experimental analysis, mitigation, recovery.
\newblock In {\em Inside Solid State Drives}. Springer, 2018.

\bibitem{cai-hpca-2017}
Yu~Cai, Saugata Ghose, Yixin Luo, Ken Mai, Onur Mutlu, and Erich~F. Haratsch.
\newblock Vulnerabilities in {MLC NAND} flash memory programming: Experimental
  analysis, exploits, and mitigation techniques.
\newblock In {\em HPCA}, 2017.

\bibitem{cai-date-2012}
Yu~Cai, Erich~F. Haratsch, Onur Mutlu, and Ken Mai.
\newblock Error patterns in {MLC NAND} flash memory: Measurement,
  characterization, and analysis.
\newblock In {\em DATE}, 2012.

\bibitem{cai-date-2013}
Yu~Cai, Erich~F. Haratsch, Onur Mutlu, and Ken Mai.
\newblock Threshold voltage distribution in {MLC NAND} flash memory:
  Characterization, analysis, and modeling.
\newblock In {\em DATE}, 2013.

\bibitem{cai-dsn-2015}
Yu~Cai, Yixin Luo, Saugata Ghose, and Onur Mutlu.
\newblock Read disturb errors in {MLC NAND} flash memory: Characterization,
  mitigation, and recovery.
\newblock In {\em DSN}, 2015.

\bibitem{cai-hpca-2015}
Yu~Cai, Yixin Luo, Erich~F. Haratsch, Ken Mai, and Onur Mutlu.
\newblock Data retention in {MLC NAND} flash memory: Characterization,
  optimization, and recovery.
\newblock In {\em HPCA}, 2015.

\bibitem{cai-iccd-2013}
Yu~Cai, Onur Mutlu, Erich~F. Haratsch, and Ken Mai.
\newblock Program interference in {MLC NAND} flash memory: Characterization,
  modeling, and mitigation.
\newblock In {\em ICCD}, 2013.

\bibitem{cai-inteltechj-2013}
Yu~Cai, Gulay Yalcin, Onur Mutlu, Erich~F. Haratsch, Adrian Crista, Osman~S.
  Unsal, and Ken Mai.
\newblock Error analysis and retention-aware management for {NAND} flash
  memory.
\newblock {\em Intel Tech. J.}, 2013.

\bibitem{cai-sigmetrics-2014}
Yu~Cai, Gulay Yalcin, Onur Mutlu, F.~Erich Haratsch, Osman Unsal, Adrian
  Cristal, and Ken Mai.
\newblock Neighbor-cell assisted error correction for {MLC NAND} flash
  memories.
\newblock In {\em SIGMETRICS}, 2014.

\bibitem{cox-fms-2018}
Alvin Cox.
\newblock {JEDEC SSD} endurance workloads.
\newblock In {\em FMS}, 2011.

\bibitem{fukami-dfrwseu-2017}
Aya Fukami, Saugata Ghose, Yixin Luo, Yu~Cai, and Onur Mutlu.
\newblock Improving the reliability of chip-off forensic analysis of {NAND}
  flash memory devices.
\newblock In {\em DFRWS EU}, 2014.

\bibitem{huh-iccss-2020}
Hwang Huh, Wanik Cho, Jinhaeng Lee, Yujong Noh, Yongsoon Park, Sunghwa Ok,
  Jongwoo Kim, Kayoung Cho, Hyunchul Lee, Geonu Kim, Kangwoo Park, Kwanho Kim,
  Heejoo Lee, Sooyeol Chai, Chankeun Kwon, Hanna Cho, Chanhui Jeong, Yujin
  Yang, Jayoon Goo, Jangwon Park, Juhyeong Lee, Heonki Kim, Kangwook Jo,
  Cheoljoong Park, Hyeonsu Nam, Hyunseok Song, Sangkyu Lee, Woopyo Jeong,
  Kun-Ok Ahn, and Tae-Sung Jung.
\newblock A {1Tb} 4b/cell 96-stacked-{WL} {3D NAND} flash memory with {30MB/s}
  program throughput using peripheral circuit under memory cell array
  technique.
\newblock In {\em ISSCC}, 2020.

\bibitem{kang-iccss-2016}
Dongku Kang, Woopyo Jeong, Chulbum Kim, Doo-Hyun Kim, Yong~Sung Cho, Kyung-Tae
  Kang, Jinho Ryu, Kyung-Min Kang, Sungyeon Lee, Wandong Kim, Hanjun Lee,
  Jaedoeg Yu, Nayoung Choi, Dong-Su Jang, Jeong-Don Ihm, Doogon Kim, Young-Sun
  Min, Moo-Sung Kim, An-Soo Park, Jae-Ick Son, In-Mo Kim, Pansuk Kwak, Bong-Kil
  Jung, Doo-Sub Lee, Hyunggon Kim, Hyang-Ja Yang, Dae-Seok Byeon, Ki-Tae Park,
  Kye-Hyun Kyung, and Jeong-Hyuk Choi.
\newblock {256Gb} 3b/cell {V-NAND} flash memory with 48 stacked {WL} layers.
\newblock In {\em ISSCC}, 2016.

\bibitem{kim-isscc-2020}
Doo-Hyun Kim, Hyunggon Kim, Sungwon Yun, Youngsun Song, Jisu Kim, Sung-Min Joe,
  Kyung-Hwa Kang, Joonsuc Jang, Hyun-Jun Yoon, Kanabin Lee, Minseok Kim,
  Joonsoo Kwon, Jonghoo Jo, Sehwan Park, Jiyoon Park, Jisoo Cho, Sohyun Park,
  Garam Kim, Jinbae Bang, Heejin Kim, Jongeun Park, Deokwoo Lee, Seonyong Lee,
  Hwajun Jang, Han-Jun Lee, Donghyun Shin, Jungmin Park, Jungkwan Kim, Jongmin
  Kim, Kichang Jang, Il~Han Park, Seung~Hyun Moon, Myung-Hoon Choi, Pansuk
  Kwak, Joo-Yong Park, Youngdon Choi, Sang-Lok Kim, Seungjae Lee, Dongku Kang,
  Jeong-Don Lim, Dae-Seok Byeon, Kiwhan Song, Junghwan Choi, Sang~Joon Hwang,
  and Jaeheon Jeong.
\newblock A {1Tb} 4b/cell {NAND} flash memory with {tPROG}=2ms, {tR}=110$\mu$s
  and 1.2{Gb/s} high-speed {IO} rate.
\newblock In {\em ISSCC}, 2020.

\bibitem{leong-uspatent-2008}
Nancy Leong, Sachit Chandra, and Hounien Chen.
\newblock Random cache read using a double memory, 2008.
\newblock {US} Patent 7,423,915.

\bibitem{luo-ieeejsac-2016}
Yixin Luo, Saugata Ghose, Yu~Cai, Erich~F. Haratsch, and Onur Mutlu.
\newblock Enabling accurate and practical online flash channel modeling for
  modern {MLC NAND} flash memory.
\newblock {\em IEEE JSAC}, 2016.

\bibitem{luo-hpca-2018}
Yixin Luo, Saugata Ghose, Yu~Cai, Erich~F. Haratsch, and Onur Mutlu.
\newblock Heatwatch: Improving {3D NAND} flash memory device reliability by
  exploiting self-recovery and temperature awareness.
\newblock In {\em HPCA}, 2018.

\bibitem{luo-sigmetrics-2018}
Yixin Luo, Saugata Ghose, Yu~Cai, Erich~F. Haratsch, and Onur Mutlu.
\newblock Improving {3D NAND} flash memory lifetime by tolerating early
  retention loss and process variation.
\newblock In {\em SIGMETRICS}, 2018.

\bibitem{macronix-technote-2013}
Macronix.
\newblock Technical note: Improving {NAND} throughput with two-plane and cache
  operations, 2013.
\newblock
  \url{https://www.macronix.com/Lists/ApplicationNote/Attachments/1907/AN0268V1_Improving\%20NAND\%20Throughput\%20with\%20Two-Plane\%20and\%20Cache\%20Operations.pdf}.

\bibitem{micron-technote-2004}
Micron.
\newblock Technical note: {NAND} flash performance increase using the {Micron}
  {PAGE READ CACHE MODE} command, 2004.
\newblock
  \url{https://www.micron.com/-/media/client/global/Documents/Products/Technical\%20Note/NAND\%20Flash/tn2901.pdf}.

\bibitem{micron-flyer-2016}
Micron.
\newblock Product flyer: {Micron 3D NAND} flash memory, 2016.
\newblock
  \url{https://www.micron.com/-/media/client/global/documents/products/product-flyer/3d_nand_flyer.pdf?la=en}.

\bibitem{shim-micro-2019}
Youngseop Shim, Myungsuk Kim, Myoungjun Chun, Jisung Park, Yoona Kim, and
  Jihong Kim.
\newblock Exploiting process similarity of {3D} flash memory for high
  performance {SSDs}.
\newblock In {\em MICRO}, 2019.

\bibitem{tavakkol-fast-2018}
Arash Tavakkol, Juan G{\'o}mez-Luna, Mohammad Sadrosadati, Saugata Ghose, and
  Onur Mutlu.
\newblock {MQSim:} a framework for enabling realistic studies of modern
  multi-queue {SSD} devices.
\newblock In {\em FAST}, 2018.

\end{thebibliography}

\end{document}